\documentstyle[aps,prl,preprint,floats,epsfig]{revtex}

\newcommand{\PLB}[3]{{Phys. Lett. B} {\bf #1}, #2 (#3)}
\newcommand{\dstlnu}{\mbox{$D^*\ell\nu$}}
\newcommand{\bdstlnu}{\mbox{$\bar{B} \rightarrow D^*\ell\nu$}}
\newcommand{\dstzlnu}{\mbox{$D^{*0}\ell^-\bar{\nu}$}}
\newcommand{\dstplnu}{\mbox{$D^{*+}\ell^-\bar{\nu}$}}
\newcommand{\dstz}{\mbox{$D^{*0}$}}
\newcommand{\dstp}{\mbox{$D^{*+}$}}
\newcommand{\dstxlnu}{\mbox{$D^* X\ell\nu$}}
\newcommand{\vcb}{\mbox{$|V_{cb}|$}}
\newcommand{\fone}{\mbox{${\cal F}(1)$}}
\newcommand{\fw}{\mbox{${\cal F}(w)$}}
\newcommand{\bbbar}{B\bar{B}}
\newcommand{\deltam}{\mbox{$\Delta m$}}
\newcommand{\cby}{\mbox{$\cos\theta_{B-D^*\ell}$}}
\newcommand{\gevoc}{\mbox{${\ \rm GeV}/c$}}
\newcommand{\mevocc}{\mbox{${\ \rm MeV}/c^2$}}
\newcommand{\invps}{\mbox{\ ${\rm ps}^{-1}$}}
\newcommand{\invfb}{\mbox{\ ${\rm fb}^{-1}$}}
\newcommand{\etal}{{\it et al.}}

\begin{document}
\preprint{\tighten\vbox{\hbox{\hfil CLNS 01/1773}
                        \hbox{\hfil CLEO 01-26}
}}

\title{\large Improved Measurement of \vcb\ Using \bdstlnu\ Decays}

\author{(CLEO Collaboration)}
\date{March 11, 2002}
\maketitle
\tighten

\begin{abstract}
We determine the weak coupling \vcb\ between the $b$ and $c$ quarks
using a sample of 3
million $B\bar{B}$ events in the CLEO detector at the Cornell Electron
Storage Ring.  
We determine
the yield of reconstructed \bdstlnu\ decays as a function of $w$, the
boost of the $D^*$ in the $B$ rest frame,
and from this we obtain the differential decay
rate $d\Gamma/dw$.  By extrapolating $d\Gamma/dw$ to $w=1$, the
kinematic end-point at which the $D^*$ is at rest relative to the $B$, we
extract the product \vcb\fone, where \fone\ is the form factor at
$w=1$.  We find $\vcb\fone =
0.0431\pm0.0013{\rm (stat.)}\pm 0.0018{\rm (syst.)}$.  

\end{abstract}
\newpage

{
\renewcommand{\thefootnote}{\fnsymbol{footnote}}

\begin{center}
R.~A.~Briere,$^{1}$ G.~P.~Chen,$^{1}$ T.~Ferguson,$^{1}$
G.~Tatishvili,$^{1}$ H.~Vogel,$^{1}$
N.~E.~Adam,$^{2}$ J.~P.~Alexander,$^{2}$ C.~Bebek,$^{2}$
B.~E.~Berger,$^{2}$ K.~Berkelman,$^{2}$
% ...
\end{center}
 
\small
\begin{center}
$^{1}${Carnegie Mellon University, Pittsburgh, Pennsylvania 15213}\\
$^{2}${Cornell University, Ithaca, New York 14853}\\
% ...
\end{center}

\setcounter{footnote}{0}
}
\pacs{12.15.Hh,13.20.He,13.25.Hw}
In the Standard Model, the weak decay of quarks is described by a
unitary $3\times 3$ matrix \cite{Ckm}.
This CKM matrix describes the flavor mixing among the quarks and the
amount of $CP$ violation through its single non-trivial phase.
Precise determinations of the CKM matrix elements are essential to
test the Standard Model.
This Letter presents an improved measurement of \vcb, the coupling of
the $b$ quark to the $c$ quark.
The CKM matrix element \vcb\ sets the length of the base of the
unitarity triangle (UT), which displays one CKM unitarity condition, 
and normalizes the constraint on
the UT from indirect $CP$ violation in $K^0$ decay.

One strategy for determining \vcb\ uses the decays 
${\bar B^0}\to\dstplnu$ and $B^-\to\dstzlnu$.
The rate for these decays, however, depends not only on \vcb\ and well-known
weak decay physics, but also on strong interaction effects,
parameterized by form factors.  
In general, these effects are difficult to quantify, but
Heavy Quark Effective Theory (HQET) offers a method for calculating
them at the kinematic point at which the final state $D^*$ is at rest
with respect to the initial $B$ meson.  
In this analysis~\cite{prd}, we take advantage of this information: we
divide the reconstructed candidates into bins of $w$, where $w$ is
the scalar product of the $B$ and $D^*$ four-velocities. Using these
yields, we
measure the differential rate $d\Gamma/dw$ for $w>1$, and
extrapolate to obtain the rate at $w=1$, which, combined with
theoretical results, gives \vcb.

We analyze 3.33 million $\bbbar$ events (3.1 \invfb)
produced on the $\Upsilon(4S)$ resonance at the Cornell Electron
Storage Ring (CESR) and detected in the CLEO II detector~\cite{cleonim}.
In addition, we use 1.6 \invfb\ of data
collected slightly below the $\Upsilon(4S)$ resonance for the purpose
of determining continuum $e^+ e^- \to q{\bar q} (q=u,d,s,c)$
backgrounds.

To identify $D^*$ candidates, we reconstruct the decay chains $D^{*+}
\to D^0
\pi^+$ and $D^{*0} \to D^0 \pi^0$ followed by $D^0\to K^-\pi^+$.  
We first combine kaon and pion candidates in hadronic events to form
$D^0$ candidates.
Signal candidates lie in the mass window $|m(K\pi)-1865| \le 20$\mevocc.
We then add a slow $\pi$ to the $D^0$ candidate to form a $D^*$.
For \dstp\ (\dstz) candidates we require $\deltam \equiv m(K\pi\pi)-
m(K\pi)$ to be within 2 \mevocc\ (3 \mevocc) of the known \dstp--$D^0$
(\dstz--$D^0$) mass difference.
For \dstp\ candidates, the $K$ and $\pi$ are fit to a common vertex, and
then the slow $\pi$ and $D$ are fit to a second vertex using a beam
spot constraint.
These constraints improve the \deltam\ resolution by about 20\%.

To $D^*$ candidates we add a lepton candidate.  
We select electrons using the ratio of the energy deposited in the CsI
calorimeter to the reconstructed track momentum, the shape of the shower
in the calorimeter, and the specific ionization in the drift
chamber.
Our candidates lie in the momentum range $0.8~<~p_e~\le~2.4$~\gevoc.
Muon candidates penetrate $\approx 5$
interaction lengths.
Only muons above about 1.4~\gevoc\ satisfy this requirement; we
therefore demand that they lie in the momentum range 
$1.4 < p_\mu \le 2.4$~\gevoc.
The charge of the lepton must match the charge of the kaon, and, for
\dstp\ candidates, be
opposite that of the slow pion.

For each candidate we compute
\begin{equation}
\cby = \frac{2E(B) E(D^*\ell) - m_B^2 - m(D^*\ell)^2}
            {2|{\bf p}(B)||{\bf p}(D^*\ell)|}.
\end{equation}
This quantity helps distinguish signal from \dstxlnu\ background
and bounds the flight direction of the $B$ relative to the $D^*$,
which is needed to calculate $w$.
We calculate $w$ for the extremes
of the $B$ flight direction, and average these two values.
The typical resolution in $w$ is 0.03.  
We divide our sample into 10
bins from 1.0 to 1.5, the final bin including candidates up to the
kinematic limit of 1.504.
For $w>1.25$, we suppress background with no loss of signal
efficiency by restricting the angle between the $D^*$ and the lepton.

At this stage, our sample of candidates contains not only \dstlnu\ 
decays, but also ${\bar B}\to D^{**}\ell\nu$ and non-resonant
${\bar B}\to D^*\pi\ell\nu$ decays (collectively referred to here as 
${\bar B}\to \dstxlnu$ decays) and various backgrounds.
In order to disentangle the \dstlnu\ from the \dstxlnu\ decays, we use
a binned maximum likelihood fit~\cite{likelihood} to the \cby\ 
distribution.
In this fit, the normalizations of the various background
distributions are fixed, and those for \dstlnu\ and \dstxlnu\ float.

The distributions of the \dstlnu\ and \dstxlnu\ decays are taken
from Monte Carlo simulation~\cite{geant}.  
Radiative ${\bar B}\to \dstlnu\gamma$ decays, modeled by
PHOTOS~\cite{photos}, are treated as signal. 
Non-resonant ${\bar B}\to D^*\pi\ell\nu$ decays are modeled using the results
of \cite{goityroberts}, and ${\bar B}\to D^{**}\ell\nu$
decays are modeled using the ISGW2~\cite{isgw2} form factors.

We account for five classes of background: continuum, combinatoric,
uncorrelated, correlated, and fake lepton.
(1)~Continuum background from $e^+e^- \to q\bar{q}$, which amounts to
about 3.5\% of the candidates in the region $-1\le \cby \le 1$ (the
``signal region''), is measured using off-resonance data taken below the
$B\bar B$ threshold.  
We normalize the \cby\ distribution to the ratio of on- to 
off-resonance luminosities, correcting
for the small energy dependence of the continuum cross section.
(2)~Combinatoric background events, those in which one or more of the
particles in the $D^*$ candidate does not come from a true $D^*$
decay, contribute 8\% (38\%) of the candidates in the
\dstp\ (\dstz) signal region.
We take the \cby\ distribution of combinatoric background events from
a \deltam\ sideband, (155, 165)~\mevocc\  for \dstp\ and (147,
165)~\mevocc\ 
for \dstz, 
and normalize using a fit to the \deltam\ distribution in each $w$ bin.
(3)~Uncorrelated background, which accounts for approximately 5\% of the
candidates in the signal region, arises when the $D^*$ and lepton come
from the decays of different $B$ mesons.  
Most of this background consists of a $D^*$ meson combined with a
secondary lepton ({\it i.e.,}  from $b\to c\to s\ell\nu$) because
primary leptons from the other $B$ have the wrong charge correlation
in the absence of
$B^0$--$\bar{B}^0$ mixing or $D^*$ production from the hadronization
of the $\bar{c}$ in $b\to c\bar{c}s$.
We obtain the uncorrelated \cby\ distribution from Monte Carlo
simulation, normalizing to the inclusive $D^*$ production rate
observed in our data in low and high $D^*$ momentum bins, 
the measured primary and secondary lepton decay
rates~\cite{roywang}, the estimated decay rate for modes in the $B\to
D^* D^{(*)} K^{(*)}$ family~\cite{ddk,pdg}, and the measured $B^0 -
\bar{B}^0$ mixing rate~\cite{pdg}.
(4)~Correlated background events are those in which the $D^*$ and lepton
are daughters of the same $B$, but the decay was not \bdstlnu\ or
${\bar B}\to\dstxlnu$.
The most common sources are ${\bar B}\to D^*\tau\nu$
followed by leptonic $\tau$ decay and $B \to D^* D_s$ followed by
semileptonic decay of the $D_s$.  
This background accounts for fewer than 0.5\% of the candidates in the
signal region and is estimated using Monte Carlo simulation.
(5)~Finally, hadrons misidentified
as leptons contribute fewer than 0.5\% of candidates in the signal region.

\begin{figure}
\begin{center}
\epsfig{file=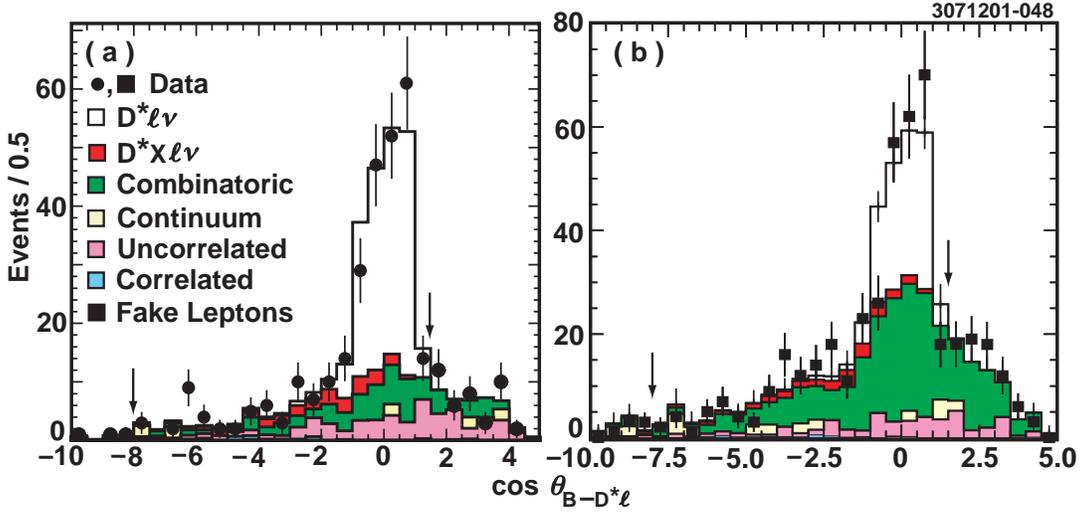,height=7cm}
\end{center}
\caption{The candidate yields for $1.10 < w < 1.15$ bin with the results of
the fit superimposed for (a) \dstplnu\ and for (b) \dstzlnu. The fit
uses the region between the arrows.
}
\label{fig:cosby}
\end{figure}

Having obtained the distributions in \cby\ of the signal and
background components, we fit for the yields 
of \dstlnu\ and \dstxlnu\ decays in
each $w$ bin.
Representative fits are shown in Fig.\ ~\ref{fig:cosby}.
The quality of the fits is good, as is agreement between the data and
fit distributions outside the fitting region.
The fit results also accurately predict the $D^*$ energy
distribution and the lepton momentum spectrum of the data.

Given the measured \dstlnu\ yields in ten bins of $w$, we fit for the
partial rate
\begin{equation}
{d\Gamma \over dw} = {G_F^2 \over 48 \pi^3} {\cal K}(w)
\left[\vcb\fw\right]^2,
\end{equation}
where ${\cal K}(w)$ is a known function of kinematic variables and
${\cal F}(w)$ is the form-factor.  
For our fit we use a form-factor parameterization~\cite{caprini}
informed by HQET and dispersion relation constraints~\cite{lebed}.
We take the form-factor ratios $R_1 (1)$ and $R_2 (1)$
from a previous measurement~\cite{ffprl} that agrees with theoretical
expectations~\cite{neubert}.
The slope $\rho^2$ of the form factor at $w=1$ is the only shape
parameter; it varies in our fit.

We fit our \dstlnu\ yields as a function of $w$ for
\vcb\fone\ and $\rho^2$.  
We minimize the sum of $\chi^2$ for \dstplnu\ and
\dstzlnu, each of which compares the measured
and expected yields in the ten reconstructed $w$ bins.
For each mode,
\begin{equation}
\chi^2 \equiv \sum_{i=1}^{10}
\frac{[N_i^{obs} - \sum_{j=1}^{10}\epsilon_{ij}N_j]^2}
{\sigma_{N_i^{obs}}^2},
\end{equation}
where $N_i^{obs}$ is the yield in the $i^{{\rm th}}$ $w$ bin,
$N_j$ is the number of decays in the $j^{{\rm th}}$ $w$ bin, and 
the matrix $\epsilon_{ij}$ accounts for the reconstruction efficiency
and the smearing in $w$.
In the above,
\begin{equation} 
N_j \equiv 4 f N_{\Upsilon (4S)}{\cal B}_{D^*} {\cal
B}_{D^0}\tau_{B}\int_{w_j}dw \frac{d\Gamma}{dw},
\end{equation}
where $\tau_{B}$ is the $B^-$ or $\bar B^0$ lifetime~\cite{pdg}, ${\cal B}_{D^*}$ is
the $D^*\to D^0\pi$ branching fraction~\cite{pdg}, ${\cal B}_{D^0}$ is
the $D^0 \to K^-\pi^+$ branching fraction~\cite{dkpi}, $N_{\Upsilon
(4S)}$ is the number of $\Upsilon (4S)$ events in the sample, and
$f$ represents $f_{00}$ or $f_{+-}$, the $\Upsilon (4S)\to B^0 \bar{B}^0$
or $B^+B^-$ branching fraction, as appropriate.  
We use the result of \cite{sylvia} for
$(f_{+-}/f_{00})(\tau_+/\tau_0)$ as a constraint in the fit, assuming
$f_{00}+f_{+-}=1$.
We assume that $B^-\to\dstzlnu$ and $\bar{B^0}\to\dstplnu$
have identical partial widths and form factors.

\begin{figure}
\centering
\epsfig{file=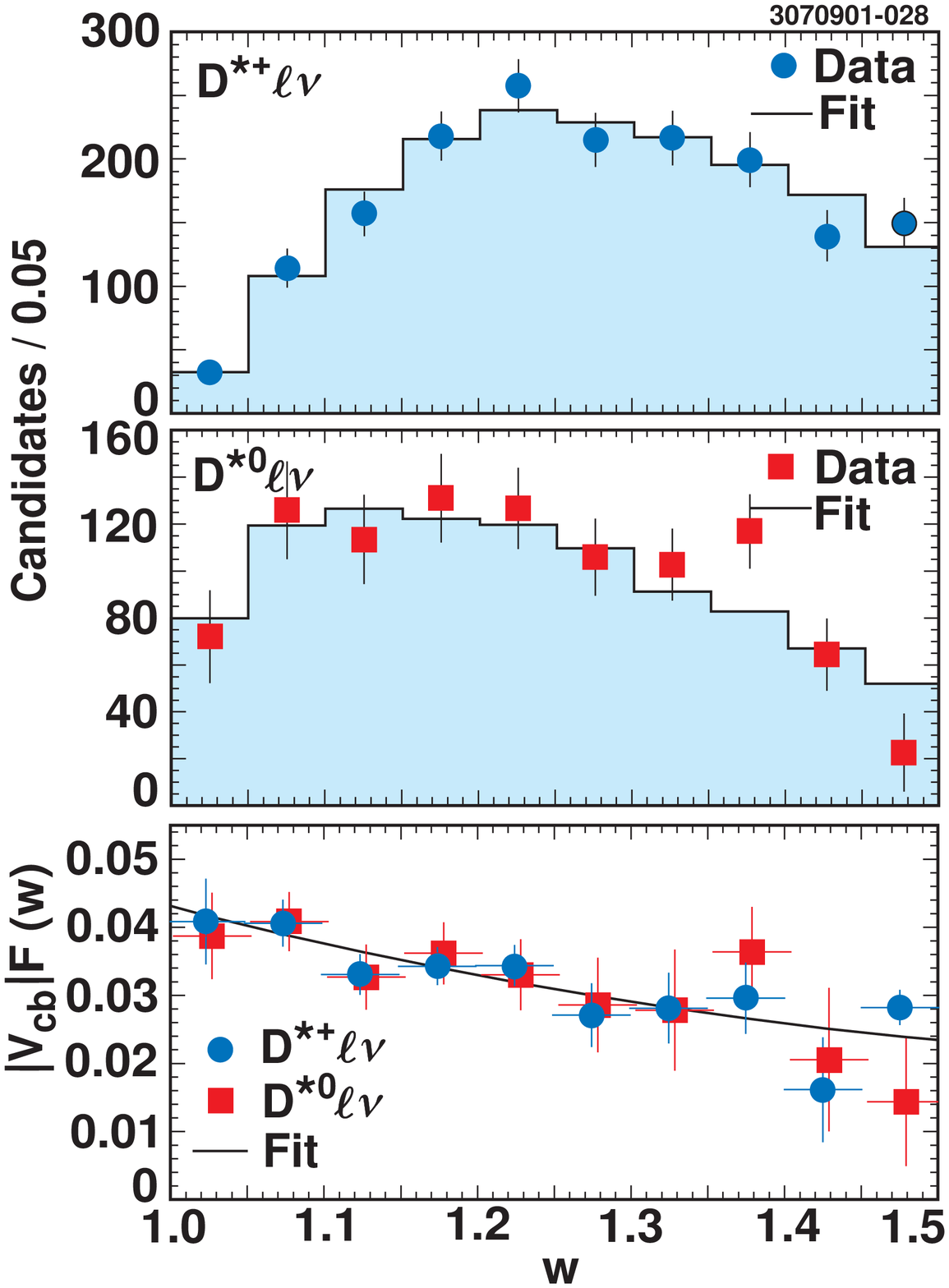,width=0.5\textwidth}
\caption{The results of the fit to the $w$ distribution.  The top
figure shows the \dstplnu\ yields (solid circles) with the results
of the fit superimposed (histogram).  The middle figure shows the same
for \dstzlnu. The bottom figure displays $\vcb\fw$,
where the solid circles (squares) are derived from the \dstplnu\ 
(\dstzlnu) yields after correcting for efficiency, smearing, and all
terms in the differential decay rate apart from \vcb\fw.  The curve
shows the result of the fit.}
\label{fig:fffit}
\end{figure}

The result of the fit is shown in Fig.\ ~\ref{fig:fffit}.  
We find
$|V_{cb}|\fone = 0.0431 \pm 0.0013 \pm 0.0018$,
$\rho^2 = 1.61 \pm 0.09 \pm 0.21$, and
$f_{+-} = 0.521 \pm 0.012$, where the errors are statistical and systematic.  
The fit $\chi^2$ is 16.8/18 d.o.f., and the
correlation coefficient between $\vcb\fone$ and $\rho^2$ is 0.86,
which becomes 0.22 after including systematic correlations.
Integrated over $w$, these parameters give $\Gamma = 0.0394\pm 0.0012
\pm 0.0026 \invps$, implying the branching fractions
${\cal B}({\bar B^0}\to \dstplnu)=(6.09\pm 0.19 \pm 0.40)$\% and 
${\cal B}({     B^-}\to \dstzlnu)=(6.50\pm 0.20 \pm 0.43)$\%.
Separate \dstplnu\ and \dstzlnu\ fits give consistent results.
When fit using the same form factor, the present data are
consistent with the previous CLEO result~\cite{oldcleo}, which
analyzed a subset of the data reported in this Letter.
The results in this Letter supersede our earlier measurement.

The systematic uncertainties are summarized in Table~\ref{tab:syserrdstp}.
The dominant systematic uncertainties arise from our background
estimations and from our knowledge of the slow-pion reconstruction
efficiency.

We test that the combinatoric candidates in the $\Delta m$ sideband have
the same \cby\ distribution as the background events in the peak region
by applying our procedure to Monte Carlo simulated events.  
The systematic uncertainty is the difference between the
results obtained using the true Monte-Carlo background and those obtained
using the sideband subtraction.
We also vary the function used to normalize the
\deltam\ sideband.
The main source of uncertainty from the uncorrelated background is the
branching fraction of the $B\to D^*D^{(*)}K^{(*)}$ 
decays, which we vary by $\pm 50\%$.  
Smaller effects arise from the primary and secondary lepton rates and
from the uncertainty in $B^0-\bar{B}^0$ mixing.
We assess the uncertainty arising from the correlated background by
varying by 50\% the branching fractions of the contributing modes.

A major source of uncertainty for the analysis is the reconstruction
efficiency of the slow pion from the $D^*$ decay.  
Due to the small energy release in $D^*$ decay, the slow-pion momentum
is correlated with $w$.
Thus, for \dstp, the efficiency is small near $w=1$ for low-momentum
pions and increases rapidly over the next few $w$ bins, while the
efficiency for $\pi^0\to\gamma\gamma$ is more uniform in $w$.
We have explored the detection efficiencies as a function of event
environment (nearby tracks or showers), drift chamber performance
(single measurement resolution and efficiency), 
vertexing requirements,
calorimeter simulation (noise, nonlinearities, and shower simulation
threshold), and description of the detector material in our simulation.
We vary the amount of noise in the calorimeter simulation and
introduce possible residual nonlinearities in the energy scale.
These variations are constrained by $m(\gamma\gamma)$ and lateral
shower shape distributions from an independent sample of $\pi^0$
candidates from our data.
The uncertainty in \vcb\fone\ is dominated by
uncertainties in the number of interaction lengths in the inner detector
(1.3\%) and the vertexing efficiency (1.5\%).

We determine the tracking efficiency uncertainties for the lepton and
the $K$ and $\pi$ forming the $D^0$ in the same study used for the
slow pion from the \dstp\ decay.  
These uncertainties are confirmed in a study of 1-prong versus 3-prong
$\tau$ decays. 
The efficiency for identifying electrons (muons) has been evaluated using
radiative Bhabha ($\mu$-pair) events embedded in hadronic events, and
has an uncertainty of 2.6\% (1.6\%).
Separate electron and muon analyses of our data give consistent results.

Finally, our analysis requires knowledge of the \cby\ distribution of
the \dstxlnu\ contribution.  
This distribution in turn depends on both the poorly known branching
fractions of contributing modes and their form factors.
We note that the ${\bar B}\to D^*\pi\ell\nu$ and ${\bar B}\to
D_1\ell\nu$ modes have the largest and smallest mean \cby. 
We therefore repeat the analysis, first using pure ${\bar B}\to
D^*\pi\ell\nu$ to describe our \dstxlnu\ decays and then using pure
${\bar B}\to D_1\ell\nu$ to describe these decays, and we take the
larger of the two excursions as our systematic error.

The form factor ratios $R_1$ and $R_2$ affect the lepton
spectrum and therefore the fraction of candidates satisfying our
lepton momentum requirements.  
To assess this effect, we vary $R_1$ and $R_2$ within their
measurement errors, taking into account their correlation.
\begin{table}
\centering
\caption{The fractional systematic uncertainties.}
\label{tab:syserrdstp}
\medskip
\begin{tabular}{lccc} 
Source  &  $|V_{cb}|\fone$(\%) &$\rho^2$(\%) & $\Gamma(B \to
\dstlnu)$(\%) \\ \hline
Continuum Background          & 0.0 & 0.2 & 0.1 \\
Combinatoric Background       & 1.6 & 2.9 & 1.3 \\
Uncorrelated Background       & 0.7 & 1.0 & 0.5 \\
Correlated Background         & 0.1 & 0.6 & 0.8 \\
Fake Leptons                  & 0.0 & 0.3 & 0.2 \\
Slow $\pi$ finding            & 2.1 & 2.8 & 2.8 \\
Vertex Reconstruction         & 1.5 & 1.6 & 2.9 \\
$K$, $\pi$,\&\ $\ell$ finding & 1.0 & 0.0 & 1.9 \\
Lepton ID                     & 0.8 & 0.6 & 1.1 \\
$B$ momentum \& mass          & 0.1 & 0.1 & 0.2 \\
\dstxlnu\ model               & 0.3 & 1.6 & 0.9 \\ 
Final-state Radiation         & 0.7 & 0.3 & 1.1 \\
Number of $B\bar{B}$ events   & 0.9 & 0.0 & 1.8 \\
$\tau_B$ and Branching Fractions
                              & 1.8 & 0.0 & 3.5 \\
$R_1(1)$ and $R_2(1)$         & 1.4 &12.0 & 1.8 \\
\hline
Total                         & 4.3 &13.0 & 6.6 \\ 
\end{tabular}
\end{table}

Using a recent lattice calculation~\cite{kronfeld} that yields
$\fone=0.919^{+0.030}_{-0.035}$, our result for \vcb\fone\ implies
\begin{equation}
\vcb =0.0469 \pm0.0014({\rm stat.}) \pm 0.0020({\rm syst.})\pm
0.0018({\rm theor.}).
\end{equation}
Our result is the most precise to date and is
somewhat higher than but marginally consistent 
with previous measurements~\cite{adob}.
However, we note that our ability to reconstruct \cby\ makes our
analysis approximately four times less sensitive to the poorly known
\dstxlnu\ background and allows us to constrain it with the data.
This value of \vcb\ is also somewhat higher than that
obtained from inclusive semileptonic $B$ decays~\cite{cleomoments}.  If
confirmed, this discrepancy could signal a violation of
quark-hadron duality.
A larger value of \vcb\ affects constraints on the CKM unitarity
triangle, reducing expectations for indirect $CP$ violation
in the $B$ system.

We gratefully acknowledge the effort of the CESR staff in providing us with
excellent luminosity and running conditions.
We thank A.~Kronfeld, W.~Marciano, and M.~Neubert for helpful discussions.
This work was supported by
the National Science Foundation,
the U.S. Department of Energy,
the Research Corporation,
and the Texas Advanced Research Program.

\nopagebreak

\end{document}